\documentclass[twocolumn,showpacs,aps,prd]{revtex4-1}
\usepackage{epsfig}
\usepackage{graphicx}
\usepackage{dcolumn}
\usepackage{bm}
\usepackage{ltablex,booktabs}
\usepackage{overpic}
\usepackage{subfigure}
\usepackage{float}
\usepackage{color}
\usepackage{amsmath}
\usepackage{mathcomp}
\usepackage{mathrsfs}
\usepackage{multirow}
\usepackage{rotating}
\usepackage{amssymb}
\usepackage{gensymb}
\usepackage{amsmath}
\usepackage{tabularx}
\usepackage{tikz-feynman}
\usepackage{lineno}
\usepackage[bookmarksnumbered, pdfstartview=FitH,colorlinks,urlcolor=blue, citecolor=blue,linkcolor=blue] {hyperref}

\begin{document}
\normalsize
\parskip=5pt plus 1pt minus 1pt
\title{\boldmath Search for the semileptonic decay $D_s^+\to \pi^0e^+\nu_e$}

\author{
\begin{small}
\begin{center}
M.~Ablikim$^{1}$, M.~N.~Achasov$^{10,b}$, P.~Adlarson$^{67}$, M.~Albrecht$^{4}$, R.~Aliberti$^{28}$, A.~Amoroso$^{66A,66C}$, M.~R.~An$^{32}$, Q.~An$^{63,50}$, X.~H.~Bai$^{58}$, Y.~Bai$^{49}$
% ...
Yi2020~Wang$^{53}$, Ying~Wang$^{51}$, Z.~Wang$^{1,50}$, Z.~Y.~Wang$^{1,55}$, Ziyi~Wang$^{55}$, D.~H.~Wei$^{12}$, F.~Weidner$^{60}$, S.~P.~Wen$^{1}$, D.~J.~White$^{59}$, U.~Wiedner$^{4}$
% ...
\\
\vspace{0.2cm}
(BESIII Collaboration)\\
\vspace{0.2cm} {\it
$^{1}$ Institute of High Energy Physics, Beijing 100049, People's Republic of China\\
% ...
$^{4}$ Bochum Ruhr-University, D-44780 Bochum, Germany\\
% ...
$^{10}$ G.I. Budker Institute of Nuclear Physics SB RAS (BINP), Novosibirsk 630090, Russia\\
% ...
$^{12}$ Guangxi Normal University, Guilin 541004, People's Republic of China\\
% ...
$^{28}$ Johannes Gutenberg University of Mainz, Johann-Joachim-Becher-Weg 45, D-55099 Mainz, Germany\\
% ...
$^{32}$ Liaoning Normal University, Dalian 116029, People's Republic of China\\
% ...
$^{49}$ Southeast University, Nanjing 211100, People's Republic of China\\
$^{50}$ State Key Laboratory of Particle Detection and Electronics, Beijing 100049, Hefei 230026, People's Republic of China\\
$^{51}$ Sun Yat-Sen University, Guangzhou 510275, People's Republic of China\\
% ...
$^{53}$ Tsinghua University, Beijing 100084, People's Republic of China\\
% ...
$^{55}$ University of Chinese Academy of Sciences, Beijing 100049, People's Republic of China\\
% ...
$^{58}$ University of Jinan, Jinan 250022, People's Republic of China\\
$^{59}$ University of Manchester, Oxford Road, Manchester, M13 9PL, United Kingdom\\
$^{60}$ University of Muenster, Wilhelm-Klemm-Str. 9, 48149 Muenster, Germany\\
% ...
$^{63}$ University of Science and Technology of China, Hefei 230026, People's Republic of China\\
% ...
$^{66}$ University of Turin and INFN, (A)University of Turin, I-10125, Turin, Italy; (B)University of Eastern Piedmont, I-15121, Alessandria, Italy; (C)INFN, I-10125, Turin, Italy\\
$^{67}$ Uppsala University, Box 516, SE-75120 Uppsala, Sweden\\
% ...
$^{b}$ Also at the Novosibirsk State University, Novosibirsk, 630090, Russia\\
% ...
}      
\end{center}
\vspace{0.4cm}
\end{small}
}
\noaffiliation{}

\begin{abstract}
  We present the first search for the semileptonic decay
  $D_s^+\to \pi^0 e^+\nu_e$ using a data sample of electron-positron
  collisions recorded with the BESIII detector at center-of-mass energies
  between 4.178 and 4.226~GeV, corresponding to an integrated luminosity of
  6.32~fb$^{-1}$. This decay is expected to be sensitive to $\pi^0$--$\eta$ mixing.
  No significant signal is observed. We set an upper limit of
  $6.4 \times 10^{-5}$ on the branching fraction at the $90\%$ confidence
  level.
\end{abstract}
\maketitle

\section{Introduction}
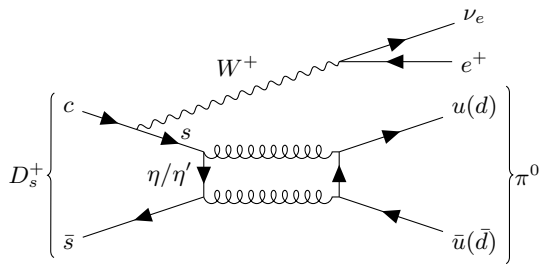
\begin{figure}
\centering
\begin{tikzpicture}[scale=0.3]
  \begin{feynman}
    \vertex (a1) at (0.0,0.0) {$c$};
    \vertex (a2) at (3.0,-1.0) ;
    \vertex (a3) at (5.2,-1.2) {$s$};
    \vertex (a4) at (6.0,-2.0) ;
    \vertex (a5) at (6.0,-4.0) ;
    \vertex (a6) at (0.0,-6.0) {$\bar{s}$};
    \vertex (a7) at (12.0,-2.0) ;
    \vertex (a8) at (12.0,-4.0) ;
    \vertex (a9) at (18.0,0.0) {$u(d)$};
    \vertex (a10) at (18.0,-6.0) {$\bar{u}(\bar{d})$};
    \vertex (a11) at (12.0,2.0) ;
    \vertex (a12) at (18.0,2.0) {$e^{+}$};
    \vertex (a13) at (18.0,4.0) {$\nu_{e}$};
    \vertex (text1) at (4.5,-3.0) {$\eta/\eta'$};
    \vertex (text2) at (7.5,1.8) {$W^{+}$};
    \diagram* {
      {[edges=fermion]
        (a1) -- (a2) -- (a4),
        (a5) -- (a6),
        (a10) -- (a8) -- (a7) -- (a9),
        (a12) -- (a11) -- (a13),
      },
      (a4) -- [fermion] (a5),
      (a4) -- [gluon] (a7),
      (a5) -- [gluon] (a8),
      (a2) -- [boson] (a11),
    };
    \draw [decoration={brace}, decorate] (a6.south west) -- (a1.north west)
          node [pos=0.5, left] {$D^{+}_{s}$};
    \draw [decoration={brace}, decorate] (a9.north east) -- (a10.south east)
          node [pos=0.5, right] {$\pi^{0}$};
  \end{feynman}
\end{tikzpicture}
\caption{Feynman diagram of the semileptonic decay $D^+_s \to \pi^0 e^+\nu_e$
  through $\pi^0$-$\eta$ mixing.}
\label{feynman}
\end{figure}

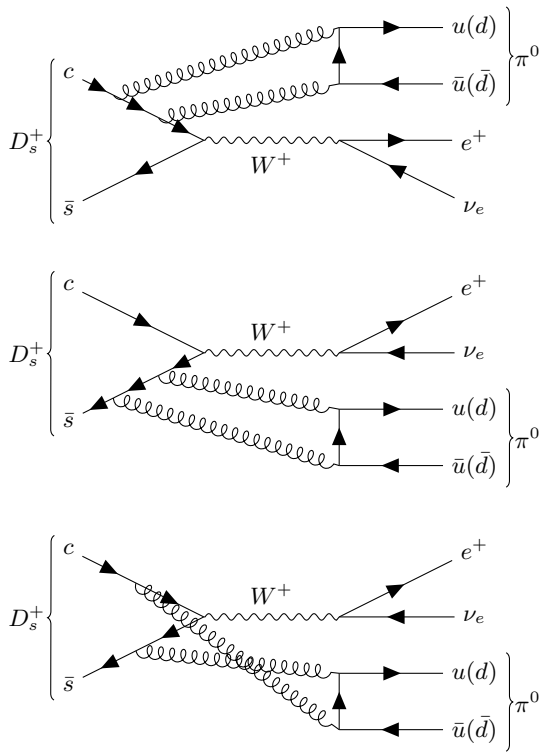
\begin{figure}
  \centering
  \begin{tikzpicture}[scale=0.3]
    \begin{feynman}
      \vertex (a1) at (0.0,0.0) {$c$};
      \vertex (a2) at (2.0,-1.0) ;
      \vertex (a3) at (4.0,-2.0) ;
      \vertex (a4) at (6.0,-3.0) ;
      \vertex (a5) at (0.0,-6.0) {$\bar{s}$};
      \vertex (a6) at (12.0,-3.0) ;
      \vertex (a7) at (18.0,2.0) {$u(d)$};
      \vertex (a8) at (18.0,-0.5) {$\bar{u}(\bar{d})$};
      \vertex (a9) at (12.0,2.0) ;
      \vertex (a10) at (12.0,-0.5) ;
      \vertex (a11) at (18.0,-3.0) {$e^{+}$};
      \vertex (a12) at (18.0,-6.0) {$\nu_{e}$};
      \vertex (text2) at (9.0,-4.0) {$W^{+}$};
      \diagram* {
        {[edges=fermion]
          (a1) -- (a2) -- (a3) -- (a4) -- (a5),
          (a12) -- (a6) -- (a11),
          (a8) -- (a10) -- (a9) -- (a7),
        },
        (a4) -- [boson] (a6),
        (a3) -- [gluon] (a10),
        (a2) -- [gluon] (a9),
      };
      \draw [decoration={brace}, decorate] (a5.south west) -- (a1.north west)
      node [pos=0.5, left] {$D^{+}_{s}$};
      \draw [decoration={brace}, decorate] (a7.north east) -- (a8.south east)
      node [pos=0.5, right] {$\pi^{0}$};
    \end{feynman}
  \end{tikzpicture}
  \begin{tikzpicture}[scale=0.3]
    \begin{feynman}
      \vertex (a1) at (0.0,0.0) {\phantom{$c$}};
      \vertex (a5) at (0.0,-0.5) {\phantom{$c$}};
      \vertex (a7) at (18.0,0) {\phantom{$c$}};
      \vertex (a8) at (18.0,-0.5) {\phantom{$c$}};
    \end{feynman}
  \end{tikzpicture}
  \begin{tikzpicture}[scale=0.3]
    \begin{feynman}
      \vertex (a1) at (0.0,0.0) {$c$};
      \vertex (a2) at (2.0,-5.0) ;
      \vertex (a3) at (4.0,-4.0) ;
      \vertex (a4) at (6.0,-3.0) ;
      \vertex (a5) at (0.0,-6.0) {$\bar{s}$};
      \vertex (a6) at (12.0,-3.0) ;
      \vertex (a7) at (18.0,-5.5) {$u(d)$};
      \vertex (a8) at (18.0,-8.0) {$\bar{u}(\bar{d})$};
      \vertex (a9) at (12.0,-5.5) ;
      \vertex (a10) at (12.0,-8.0) ;
      \vertex (a11) at (18.0,0.0) {$e^{+}$};
      \vertex (a12) at (18.0,-3.0) {$\nu_{e}$};
      \vertex (text2) at (9.0,-2.0) {$W^{+}$};
      \diagram* {
        {[edges=fermion]
          (a1) -- (a4) -- (a3) -- (a2) -- (a5) ,
          (a12) -- (a6) -- (a11),
          (a8) -- (a10) -- (a9) -- (a7),
        },
        (a4) -- [boson] (a6),
        (a3) -- [gluon] (a9),
        (a2) -- [gluon] (a10),
      };
      \draw [decoration={brace}, decorate] (a5.south west) -- (a1.north west)
      node [pos=0.5, left] {$D^{+}_{s}$};
      \draw [decoration={brace}, decorate] (a7.north east) -- (a8.south east)
      node [pos=0.5, right] {$\pi^{0}$};
    \end{feynman}
  \end{tikzpicture}
  \begin{tikzpicture}[scale=0.3]
    \begin{feynman}
      \vertex (a1) at (0.0,0.0) {\phantom{$c$}};
      \vertex (a5) at (0.0,-0.5) {\phantom{$c$}};
      \vertex (a7) at (18.0,0) {\phantom{$c$}};
      \vertex (a8) at (18.0,-0.5) {\phantom{$c$}};
    \end{feynman}
  \end{tikzpicture}
  \begin{tikzpicture}[scale=0.3]
    \begin{feynman}
      \vertex (a1) at (0.0,0.0) {$c$};
      \vertex (a2) at (3.0,-1.5) ;
      \vertex (a3) at (3.0,-4.5) ;
      \vertex (a4) at (6.0,-3.0) ;
      \vertex (a5) at (0.0,-6.0) {$\bar{s}$};
      \vertex (a6) at (12.0,-3.0) ;
      \vertex (a7) at (18.0,-5.5) {$u(d)$};
      \vertex (a8) at (18.0,-8.0) {$\bar{u}(\bar{d})$};
      \vertex (a9) at (12.0,-5.5) ;
      \vertex (a10) at (12.0,-8.0) ;
      \vertex (a11) at (18.0,0.0) {$e^{+}$};
      \vertex (a12) at (18.0,-3.0) {$\nu_{e}$};
      \vertex (text2) at (9.0,-2.0) {$W^{+}$};
      \diagram* {
        {[edges=fermion]
          (a1) -- (a2) -- (a4) -- (a3) -- (a5) ,
          (a12) -- (a6) -- (a11),
          (a8) -- (a10) -- (a9) -- (a7),
        },
        (a4) -- [boson] (a6),
        (a3) -- [gluon] (a9),
        (a2) -- [gluon] (a10),
      };
      \draw [decoration={brace}, decorate] (a5.south west) -- (a1.north west)
      node [pos=0.5, left] {$D^{+}_{s}$};
      \draw [decoration={brace}, decorate] (a7.north east) -- (a8.south east)
      node [pos=0.5, right] {$\pi^{0}$};
    \end{feynman}
  \end{tikzpicture}
  \caption{Feynman diagrams of the semileptonic decay $D^+_s \to \pi^0
    e^+\nu_e$ through the weak annihilation effect with the
    radiation of a $\pi^0$ meson.
  }
  \label{feynman2}
\end{figure}

Neutral mesons that have hidden flavors and the same quantum
numbers can mix via the strong and electromagnetic
interactions. Meson mixing is an interesting phenomenon that can be
used to explain some specific decay processes of heavy mesons. Many
mixing effects are being widely studied, such as in the systems
$\pi^0$-$\eta$~\cite{plb-811-135879}, $\rho-\omega$~\cite{prl-80-1834}, $\omega$-$\phi$~\cite{prd-79-074006}, and
$\eta$-$\eta^{\prime}$~\cite{prd-86-117505}. This analysis
searches for $\pi^0$-$\eta$ mixing in semileptonic $D_s^{+}$
decays. The semileptonic decay $D_s^+\to\pi^0 e^+\nu_e$ can only occur
via $\pi^0$-$\eta$ mixing, as shown in Fig.~\ref{feynman}, and
nonperturbative weak annihilation effects, as shown in
Fig.~\ref{feynman2}, where the two gluons can be emitted from the $c$
quark or $\bar{s}$ quark, or one gluon from each
quark~\cite{plb-811-135879}.  However, the radiation of a $\pi^0$ from
the weak annihilation effect is suppressed not only by the
Okubo-Zweig-Iizuka~(OZI) rule but also by isospin conservation.
Consequently, the weak annihilation contribution to the
$D_s^+\to\pi^0 e^+\nu_e$ decay is relatively small compared to that
from $\pi^0$-$\eta$ mixing. The contribution to the branching
fraction~(BF) of $D_s^+\to\pi^0 e^+\nu_e$ from the weak annihilation
effect is expected to be only of the order of $10^{-7}-10^{-8}$, while the
contribution from $\pi^0$-$\eta$ mixing is expected to be
$(2.65\pm0.38)\times 10^{-5}$~\cite{plb-811-135879}. Therefore, this
decay provides an excellent opportunity to study the $\pi^0$-$\eta$
mixing effect.

In this paper, we present the first search for the semileptonic decay $D_s^+\to \pi^0 e^+\nu_e$
in a data sample corresponding to an
integrated luminosity of $6.32~\mathrm{fb}^{-1}$, which was recorded by
the BESIII detector at center-of-mass (CM) energies ($\sqrt{s}$) between
4.178 and 4.226~GeV. A blind analysis is performed to avoid possible
bias. The signal region of the data sample is only uncovered after the
event selection and analysis strategy are studied and verified based
on an ensemble of forty inclusive MC samples with the same size as the data sample.
Throughout this paper, charge conjugate channels are implied.

\section{DETECTOR and DATA SETS} \label{sec:detector_dataset}
The BESIII detector~\cite{Ablikim:2009aa, Ablikim:2019hff} records symmetric
$e^+e^-$ collisions provided by the BEPCII storage
ring~\cite{Yu:IPAC2016-TUYA01}, which operates in the CM energy range from 2.0
to 4.9~GeV. The cylindrical core of the BESIII detector covers 93\% of the full
solid angle and consists of a helium-based multilayer drift chamber~(MDC), a
plastic scintillator time-of-flight system~(TOF), and a CsI(Tl) electromagnetic
calorimeter~(EMC), which are all enclosed in a superconducting solenoidal
magnet providing a 1.0~T magnetic field. The solenoid is supported by an
octagonal flux-return yoke with resistive plate counter muon identification
modules interleaved with steel. The charged-particle momentum resolution at
$1~{\rm GeV}/c$ is $0.5\%$, and the specific energy loss (${\rm
  d}E/{\rm d}x$)
resolution is $6\%$ for
electrons from Bhabha scattering. The EMC measures photon energies with a
resolution of $2.5\%$ ($5\%$) at $1$~GeV in the barrel (end cap) region. The
time resolution in the TOF barrel region is 68~ps, while that in the end cap
region is 110~ps. The end cap TOF system was upgraded in 2015 using multi-gap
resistive plate chamber technology, providing a time resolution of
60~ps~\cite{etof}.

The data samples used in this analysis correspond to an integrated
luminosity~($\mathcal{L}_{\rm int}$) of 6.32~fb$^{-1}$ taken in the
range of $\sqrt{s} =  4.178$ to 4.226~GeV, as listed in Table~\ref{energe}.
All data samples except the 4.226~GeV one benefit from the improved time
resolution in the endcaps. In these energies, $D_{s}^{*\pm}D_s^{\mp}$ events 
provide a large sample of $D_{s}^{\pm}$ mesons.
The cross section of $D_{s}^{*\pm}D_{s}^{\mp}$ production in
$e^{+}e^{-}$ annihilation is about a factor of twenty larger than that of
$D_{s}^{+}D_{s}^{-}$~\cite{DsStrDs}, and $D_{s}^{*\pm}$ decays to
$\gamma D_{s}^{\pm}$ with a dominant BF of $(93.5\pm0.7)$\%~\cite{PDG}. 
Therefore, we use $D^{*\pm}_sD^{\mp}_s\to \gamma D_{s}^{\pm}D_{s}^{-}$ events
in this analysis.

\begin{table}[htb]
 \renewcommand\arraystretch{1.25}
 \centering
 \caption{Integrated luminosity $\mathcal{L}_{\rm int}$ and the
   recoil mass $M_{\rm rec}$ requirements for various energies, where
   $M_{\rm rec}$ is defined in Eq.~(\ref{eq:mrec}). The first and
   second uncertainties are statistical and systematic, respectively. The data collected at $\sqrt{s} = $ 4.178-4.219 GeV (which corresponds to about 83.3\% of total data sample) use the updated TOF~\cite{BESIII:2022xii, BESIII:2015zbz}.}
 \begin{tabular}{ccc}
 \hline 
 $\sqrt{s}$ (GeV) & $\mathcal{L}_{\rm int}$ (pb$^{-1}$) & $M_{\rm rec}$ (GeV/$c^2$)\\
 \hline
  4.178 &$3189.0\pm0.2\pm31.9$&[2.050, 2.180] \\
  4.189 &$526.7\pm0.1\pm2.2$&[2.048, 2.190] \\
  4.199 &$526.0\pm0.1\pm2.1$&[2.046, 2.200] \\
  4.209 &$517.1\pm0.1\pm1.8$&[2.044, 2.210] \\
  4.219 &$514.6\pm0.1\pm1.8$&[2.042, 2.220] \\
  4.226 &$1056.4\pm0.1\pm7.0$&[2.040, 2.220] \\
  \hline
 \end{tabular}
 \label{energe}
\end{table}

Large samples of Monte Carlo~(MC) simulated events produced with
{\sc geant4}-based~\cite{GEANT4} software, which includes the geometric
description of the BESIII detector and the detector response, are used to
determine the detection efficiency and to estimate the background
contributions. The simulation includes the beam-energy spread and initial-state
radiation~(ISR) in the $e^+e^-$ annihilation modeled with the generator
{\sc kkmc}~\cite{KKMC}. Inclusive MC samples with 40 times the size of data are
used to simulate the background contributions.  The inclusive MC samples, which
contain no signal $D_s^+ \to \pi^0 e^{+} \nu_{e}$ decays, include the
production of open-charm processes, the ISR production of vector
charmonium(-like) states, and the continuum processes incorporated in
{\sc kkmc}. The known decay modes are modeled with {\sc evtgen}~\cite{EVTGEN}
using world averaged BF values~\cite{PDG}, and the remaining unknown decays
from the charmonium states with {\sc lundcharm}~\cite{LUNDCHARM}. Final-state
radiation from charged final-state particles is incorporated with
{\sc photos}~\cite{PHOTOS}. The signal detection efficiencies and signal shapes
are obtained from signal MC samples, in which the signal
$D_s^+ \to \pi^0 e^{+}\nu_{e}$ decay is simulated using the {\sc ISGW2}
model~\cite{ISGW, ISGW2}.

\section{DATA ANALYSIS}
\label{chap:event_selection}
The process
$e^{+}e^{-} \to D_{s}^{*+}D_{s}^{-}+c.c.\to \gamma D_{s}^{+}D_{s}^{-}$
allows the study of semileptonic $D_{s}^{+}$ decays with a tag
technique~\cite{MarkIII-tag} since only one neutrino escapes undetected. There
are two types of samples used in the tag technique: single tag (ST) and double
tag (DT) events. In the ST sample, a $D_{s}^{-}$ meson is reconstructed through a
specific hadronic decay without any requirement on the remaining measured
tracks and EMC showers. In the DT sample, a $D_{s}^{-}$, designated as the
``tag'', is reconstructed through a hadronic decay mode first, and then the
decay $D_{s}^{+}\to \pi^0e^+\nu_e$, designated as the ``signal'', is
reconstructed with the remaining tracks and EMC showers. For a specific tag
mode, the ST yield is given by
\begin{eqnarray}
  \begin{aligned}
    N_{\text{tag}}^{\text{ST}} = 2N_{D_s^*D_s}\mathcal{B}_{\text{tag}}\epsilon_{\text{tag}}^{\text{ST}}\,, \label{eq-ST}
    \label{eq:Ntag}
  \end{aligned}
\end{eqnarray}
and the DT yield is given by
\begin{eqnarray}
  \begin{aligned}
    N_{\text{tag,sig}}^{\text{DT}}=2N_{D_s^*D_s}\mathcal{B}_{\gamma}\mathcal{B}_{\pi^0}\mathcal{B}_{\text{tag}}\mathcal{B}_{\text{sig}}\epsilon_{\text{tag,sig}}^{\text{DT}}\,,
    \label{eq:Nsig}
  \end{aligned}
\end{eqnarray}
where $N_{ D_s^*D_s }$ is the total number of $D_s^{*+}D_s^{-}+c.c.$ pairs
produced, $\mathcal{B}_{\rm sig (tag)}$ is the BF of the signal decay~(the tag
mode), $\mathcal{B}_{\gamma(\pi^0)}$ is the BF of
$D_s^* \to\gamma D_s$~($\pi^0\to\gamma\gamma$), and
$\epsilon_{\rm tag}^{\rm ST}$ $(\epsilon_{\rm tag,sig}^{\rm DT})$ is the
corresponding ST~(DT) efficiency. By isolating $\mathcal{B}_{\text{sig}}$, one
obtains
\begin{eqnarray}
  \begin{aligned}
    \mathcal{B}_{\text{sig}}=\frac{N_{\text{tag,sig}}^{\text{DT}}\epsilon_{\text{tag}}^{\text{ST}}}{\mathcal{B}_{\gamma}\mathcal{B}_{\pi^0}
      N_{\text{tag}}^{\text{ST}}\epsilon_{\text{tag,sig}}^{\text{DT}}},\,
    \label{eq:Bsig}
  \end{aligned}
\end{eqnarray}
where the yields $N_{\text{tag}}^{\text{ST}}$ and
$N_{\text{tag,sig}}^{\text{DT}}$ are obtained from data samples, while
$\epsilon_{\text{tag}}^{\text{ST}}$ and
$\epsilon_{\text{tag,sig}}^{\text{DT}}$ are obtained from inclusive and
signal MC samples, respectively. For
multiple tag modes and energy points, the above equation is generalized as 
\begin{eqnarray}
  \begin{aligned}
    \mathcal{B}_{\text{sig}}=\frac{N_{\text{total,sig}}^{\text{DT}}}{\mathcal{B}_{\gamma}\mathcal{B}_{\pi^0}\sum_{\alpha,
        i} N_{\alpha, i}^{\text{ST}}\epsilon^{\text{DT}}_{\alpha,\text{sig},
        i}/\epsilon_{\alpha, i}^{\text{ST}}},\, \label{eq:Bsig-gen}
  \end{aligned}
\end{eqnarray}
where $\alpha$ represents tag modes, $i$ represents different energy points,
and $N_{\text{total,sig}}^{\text{DT}}$ is the total signal yield.

The tag candidates are reconstructed with $K^\pm$, $\pi^\pm$, $\pi^0$,
$\rho^{0}$, $\eta$, $\eta^{\prime}$, and $K^0_S$ mesons that satisfy the
particle selection criteria detailed below. Twelve tag modes are used, and the
requirements on the invariant masses of tagged $D_s^{-}$
candidates~($M_{\rm tag}$) are summarized in Table~\ref{tab:tag-eff}.

Photon candidates are reconstructed from isolated clusters found in the
EMC. The EMC shower time is required to be within [0, 700]~ns from the
event start time in order to suppress fake photons due to electronic
noise or $e^+e^-$ beam background. Photon candidates within
$\vert\!\cos\theta\vert <0.80$ (barrel) are required to deposit more
than 25~MeV of energy, and those with
$0.86<\vert\!\cos\theta\vert<0.92$ (end cap) must deposit more than
50~MeV, where $\theta$ is the polar angle with respect to the $z$
direction (the positive direction of the MDC axis).  To exclude
showers that originate from charged tracks, the angle between the
position of each shower in the EMC and the closest extrapolated
charged track must be greater than 10 degrees.  The $\pi^0$ $(\eta)$
candidates are reconstructed through $\pi^0\to \gamma\gamma$ ($\eta
\to \gamma\gamma$) decays, with at least one barrel photon. The
diphoton invariant masses for the identification of $\pi^{0}$ and
$\eta$ decays are required to be in the ranges of $[0.115,
  0.150]$~GeV/$c^{2}$ and $[0.500, 0.570]$~GeV/$c^{2}$, respectively.
The $\chi^{2}$ of the kinematic fit constraining $M_{\gamma\gamma}$ to
the $\pi^{0}$ or $\eta$ known mass~\cite{PDG} is required to be less
than 30.

Charged particle candidates reconstructed using the information of the
MDC must satisfy $\vert\!\cos\theta\vert<0.93$ with the distance of closest approach
to the interaction point (IP) less than 10~cm in the $z$ direction and less
than 1~cm in the plane perpendicular to $z$.
Particle identification~(PID) of charged kaons and pions is implemented by
combining the information of ${\rm d}E/{\rm d}x$ from the MDC and the time-of-flight from
the TOF system. For charged kaon (pion) candidates, the likelihood for the
kaon (pion) hypothesis is required to be larger than that for a pion (kaon).
Electron PID uses EMC information along with ${\rm d}E/{\rm d}x$ and time-of-flight
to construct likelihoods for electron, pion, and kaon hypotheses
(${\cal L}_{e}, {\cal L}_{\pi}$, and ${\cal L}_{K}$).  Electron candidates must
satisfy ${\cal L}_{e}/({\cal L}_{e}+{\cal L}_{\pi}+{\cal L}_{K})>0.8$.
Additionally, the energy deposited in the EMC by the electron
candidate must be more than $80\%$ of the track momentum measured by the MDC.

Candidate $K_{S}^{0}$ mesons are reconstructed with pairs of two
oppositely charged particles, whose distances of closest approach
to the IP along $z$ are less than 20~cm. These two particles are assumed to be
pions without PID applied. Primary and secondary vertices are
reconstructed, and the decay length between the two vertices is
required to be greater than twice its uncertainty.  This requirement
is not applied for the $D_s^-\to K_S^0K^-$ decay due to the low
combinatorial background. Candidate $K_{S}^{0}$ mesons are required to
have the $\chi^2$ of the vertex fit less than 100 and be inside an
invariant-mass window $[0.487, 0.511]$~GeV/$c^{2}$, which is about
three times the resolution. The invariant mass of the $\pi^+\pi^-$
pair of the $D_s^-\to K^-\pi^+\pi^-$ decay is required to be outside
of the $K_{S}^{0}$ invariant mass window to prevent an event being
doubly counted in selecting the $D_s^-\to K_S^0K^-$ and $D_s^-\to
K^-\pi^+\pi^-$ tag modes. The $\rho^{0}$ candidates are selected via
the process $\rho^{0} \to \pi^{+}\pi^{-}$ with an invariant mass
window $[0.620, 0.920]$~GeV/$c^2$, which is about two times the
$\rho^{0}$ width. The $\eta^{\prime}$ candidates are formed from
$\pi^{+}\pi^{-}\eta$ and $\gamma\rho^{0}$ combinations with invariant
masses falling within the range of $[0.946, 0.970]$~GeV/$c^{2}$, about
three times the resolution.

In order to identify the process $e^+e^-\to D^{*\pm}_sD^{\mp}_s$, the signal
windows, listed in Table~\ref{energe}, are applied to the recoiling
mass~($M_{\rm rec}$) of the tag candidate. The definition of $M_{\rm rec}c^2$ is
\begin{eqnarray}
\begin{aligned}
  \sqrt{\left(E_{\rm cm}-\sqrt{c^2|\vec{p}_{\rm tag}|^2+c^4m^2_{D_s}}\right)^2-c^2|\vec{p}_{\rm tag}|^{2}}\,, 
\label{eq:mrec}
\end{aligned} \end{eqnarray} 
where $E_{\rm cm}$ is the energy of the $e^+e^-$ CM system,
$(\sqrt{|\vec{p}_{\rm tag}|^2+c^2m^2_{D_s}},\, \vec{p}_{\rm tag})\equiv p_{\rm tag}$
is the measured four-momentum of the tag candidate, and $m_{D_s}$ is the
known $D^{-}_{s}$ mass~\cite{PDG}. If there are multiple candidates for a tag
mode, the one with $M_{\rm rec}$ closest to the known $D_s^{*\pm}$ mass~\cite{PDG} is
chosen.

\begin{figure*}[htp]
  \begin{center}
    \includegraphics[width=0.7\textwidth]{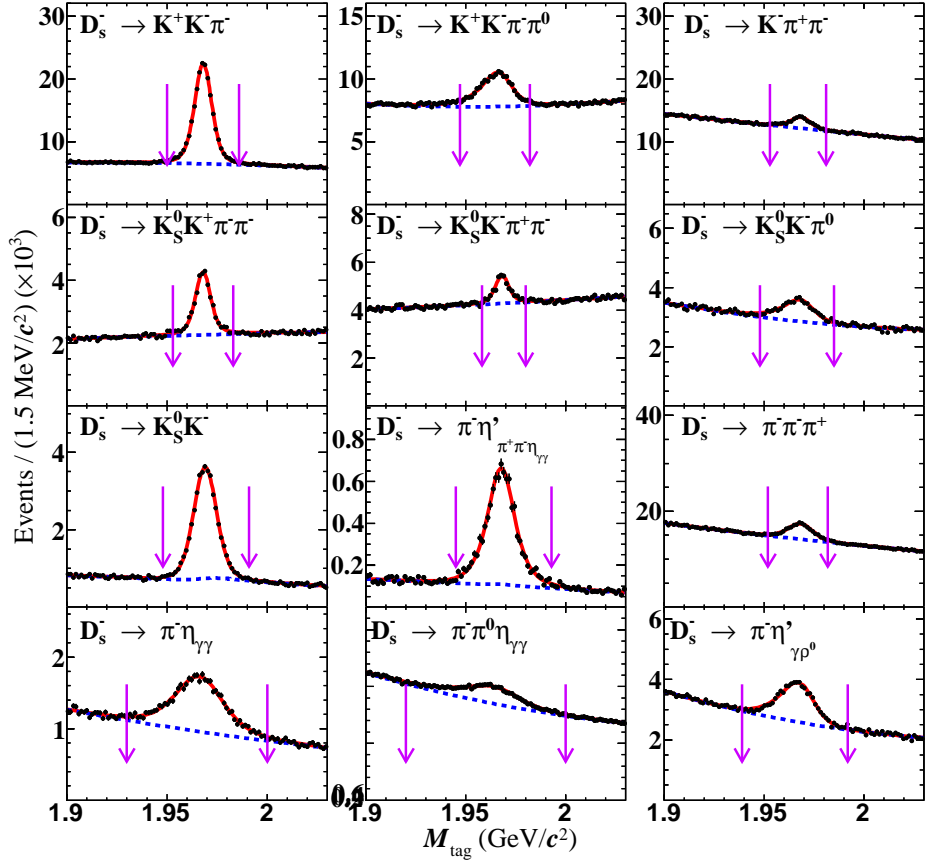}
    \caption{Fits to the $M_{\rm tag}$ distributions of the ST $D^-_s$
      candidates at $\sqrt{s} = 4.178$~GeV. The points with error bars
      are data, red solid lines are total fits, and blue dashed lines
      are the fitted backgrounds. The pairs of pink arrows denote
      signal regions. The peaking background MC-simulated shapes of
      $D^{-} \to K_{S}^{0} \pi^-$ and $D_{s}^{-} \to
      \eta\pi^+\pi^-\pi^-$ decays are added to the background
      polynomials in the fits of $D_{s}^{-} \to K_{S}^{0}K^{-}$ and
      $D_{s}^{-} \to \pi^{-}\eta^{\prime}$ decays to account for the
      peaking background, respectively.  }
    \label{fit:Mass-data-Ds_4180}
  \end{center}
\end{figure*}

The ST yields for various tag modes $N_{\text{tag}}^{\text{ST}}$ are
obtained by fitting the $M_{\rm tag}$ distributions of the accepted ST
$D^-_s$ candidates. Example fits to the data sample at 4.178~GeV are
shown in Fig.~\ref{fit:Mass-data-Ds_4180}. The description of the
signal shape is based on the MC-simulated shape convolved with a
Gaussian function accounting for the resolution difference between
data and MC. The background is described by a second-order Chebyshev
polynomial. The only two significant peaking backgrounds in all the
tag modes are from $D^{-} \to K_{S}^{0} \pi^-$ and $D_{s}^{-} \to
\eta\pi^+\pi^-\pi^-$ decays faking the $D_{s}^{-} \to K_{S}^{0} K^-$
and $D_{s}^{-} \to \pi^-\eta^{\prime}$ tag modes, respectively. The
$D^{-} \to K_{S}^{0} \pi^-$ and $D_{s}^{-} \to \eta\pi^+\pi^-\pi^-$
background contributions are estimated to be $1724\pm34$ and $89\pm5$ events
according to the BFs given by Refs.~\cite{PDG}
and~\cite{prd-104-L071101}, which correspond to about 0.3\% and less
than 0.1\% of the total ST yields, respectively. For these cases, the
sizes and MC-simulated shapes of the two peaking backgrounds are fixed
based on their BFs and added to the background polynomials. The ST
yields in data and ST efficiencies for various tag modes are listed in
Table~\ref{tab:tag-eff}.

\begin{table*}[htbp]
  \renewcommand\arraystretch{1.25}
  \caption{Requirements on $M_{\rm tag}$, the ST yields~($N_{\text{tag}}^{\text{ST}}$) 
    and ST efficiencies ($\epsilon_{\text{tag}}^{\text{ST}}$) at $\sqrt{s}=$ (I)
    4.178~GeV, (II) 4.189-4.219~GeV, and (III) 4.226~GeV,
    where the subscripts of $\eta$ and $\eta^\prime$ denote the decay modes used
    to reconstruct $\eta$ and $\eta^\prime$ candidates. The efficiencies for the energy
    points $4.189$-$4.219$~GeV are averaged based on the luminosities. The BFs
    of the sub-particle ($K_{S}^{0}$, $\pi^{0}$, $\eta$ and $\eta^{'}$) decays
    are not included. Uncertainties are statistical only.}\label{tab:tag-eff}
  \begin{center}
    \begin{tabular}{lccccccc}
      \hline
      Tag mode & $M_{\rm tag}$ (GeV/$c^{2}$)  & (I) $N_{\rm tag}^{\rm ST}$  & (I) $\epsilon_{\rm tag}^{\rm ST} (\%)$ & (II) $N_{\rm tag}^{\rm ST}$ & (II) $\epsilon_{\rm tag}^{\rm ST} (\%)$ & (III) $N_{\rm tag}^{\rm ST}$ & (III) $\epsilon_{\rm tag}^{\rm ST} (\%)$\\
      \hline
      $D_{s}^{-} \to K_{S}^{0}K^{-}$
      & [1.948, 1.991] & $\phantom{0}31941 \pm 312$   & $47.36 \pm 0.07$ & $18559            \pm 261$ & $47.26 \pm 0.09$ & $\phantom{0}6582 \pm 160$  & $46.37 \pm 0.16$\\
      $D_{s}^{-} \to K^{+}K^{-}\pi^{-}$
      & [1.950, 1.986] & $137240           \pm 614$   & $39.47 \pm 0.03$ & $81286            \pm 505$ & $39.32 \pm 0.04$ & $28439           \pm 327$  & $38.38 \pm 0.07$\\
      $D_{s}^{-} \to K_{S}^{0}K^{-}\pi^{0}$
      & [1.946, 1.987] & $\phantom{0}11385 \pm 529$   & $16.12 \pm 0.11$ & $\phantom{0}6832  \pm 457$ & $15.71 \pm 0.16$ & $\phantom{0}2227 \pm 220$  & $15.93 \pm 0.29$\\
      $D_{s}^{-} \to K^{+}K^{-}\pi^{-}\pi^{0}$
      & [1.947, 1.982] & $\phantom{0}39306 \pm 799$   & $10.50 \pm 0.03$ & $23311            \pm 659$ & $10.58 \pm 0.05$ & $\phantom{0}7785 \pm 453$  & $10.39 \pm 0.08$\\
      $D_{s}^{-} \to K_{S}^{0}K^{-}\pi^{-}\pi^{+}$
      & [1.958, 1.980] & $\phantom{00}8093 \pm 326$   & $20.40 \pm 0.12$ & $\phantom{0}5269  \pm 282$ & $20.19 \pm 0.17$ & $\phantom{0}1662 \pm 217$  & $19.50 \pm 0.31$\\
      $D_{s}^{-} \to K_{S}^{0}K^{+}\pi^{-}\pi^{-}$
      & [1.953, 1.983] & $\phantom{0}15719 \pm 289$   & $21.83 \pm 0.06$ & $\phantom{0}8948  \pm 231$ & $21.63 \pm 0.09$ & $\phantom{0}3263 \pm 172$  & $21.29 \pm 0.15$\\
      $D_{s}^{-} \to \pi^{-}\pi^{-}\pi^{+}$
      & [1.952, 1.982] & $\phantom{0}37977 \pm 859$   & $51.43 \pm 0.15$ & $21909            \pm 776$ & $50.35 \pm 0.22$ & $\phantom{0}7511 \pm 393$  & $49.32 \pm 0.41$\\
      $D_{s}^{-} \to \pi^{-}\eta_{\gamma\gamma}$
      & [1.930, 2.000] & $\phantom{0}17940 \pm 403$   & $43.58 \pm 0.15$ & $10025            \pm 339$ & $43.00 \pm 0.22$ & $\phantom{0}3725 \pm 252$  & $41.83 \pm 0.41$\\
      $D_{s}^{-} \to \pi^-\pi^0\eta$
      & [1.920, 2.000] & $\phantom{00}42618 \pm 1397$ & $18.09 \pm 0.11$ &$\phantom{0}26067 \pm 1196$ & $18.40 \pm 0.16$ & $\phantom{0}10513 \pm 1920$& $17.69 \pm 0.30$\\
      $D_{s}^{-} \to \pi^{-}\eta_{\pi^{+}\pi^{-}\eta_{\gamma\gamma}}^{'}$
      & [1.940, 1.996] & $\phantom{00}7759 \pm 141$ & $19.12 \pm 0.06$ & $\phantom{0}4428 \pm 111$ & $19.00 \pm 0.08$ &$\phantom{0}1648 \pm 74\phantom{0}$ & $18.56 \pm 0.13$\\
      $D_{s}^{-} \to \pi^{-}\eta^{\prime}_{\gamma\rho^0}$
      & [1.939, 1.992] & $\phantom{0}20610 \pm 538$   & $26.28 \pm 0.10$ & $11937 \pm 480$            & $26.09 \pm 0.14$ & $\phantom{0}3813 \pm 335$  & $25.94 \pm 0.27$\\
      $D_{s}^{-} \to K^{-}\pi^{+}\pi^{-}$
      & [1.953, 1.986] & $\phantom{0}17423 \pm 666$   & $47.46 \pm 0.22$ & $10175           \pm 448$  & $47.19 \pm 0.32$ & $\phantom{0}4984 \pm 458$  & $45.66 \pm 0.59$\\
      \hline
    \end{tabular}
  \end{center}
\end{table*}

\begin{figure*}[htp]
  \begin{center}
    \includegraphics[width=0.32\textwidth]{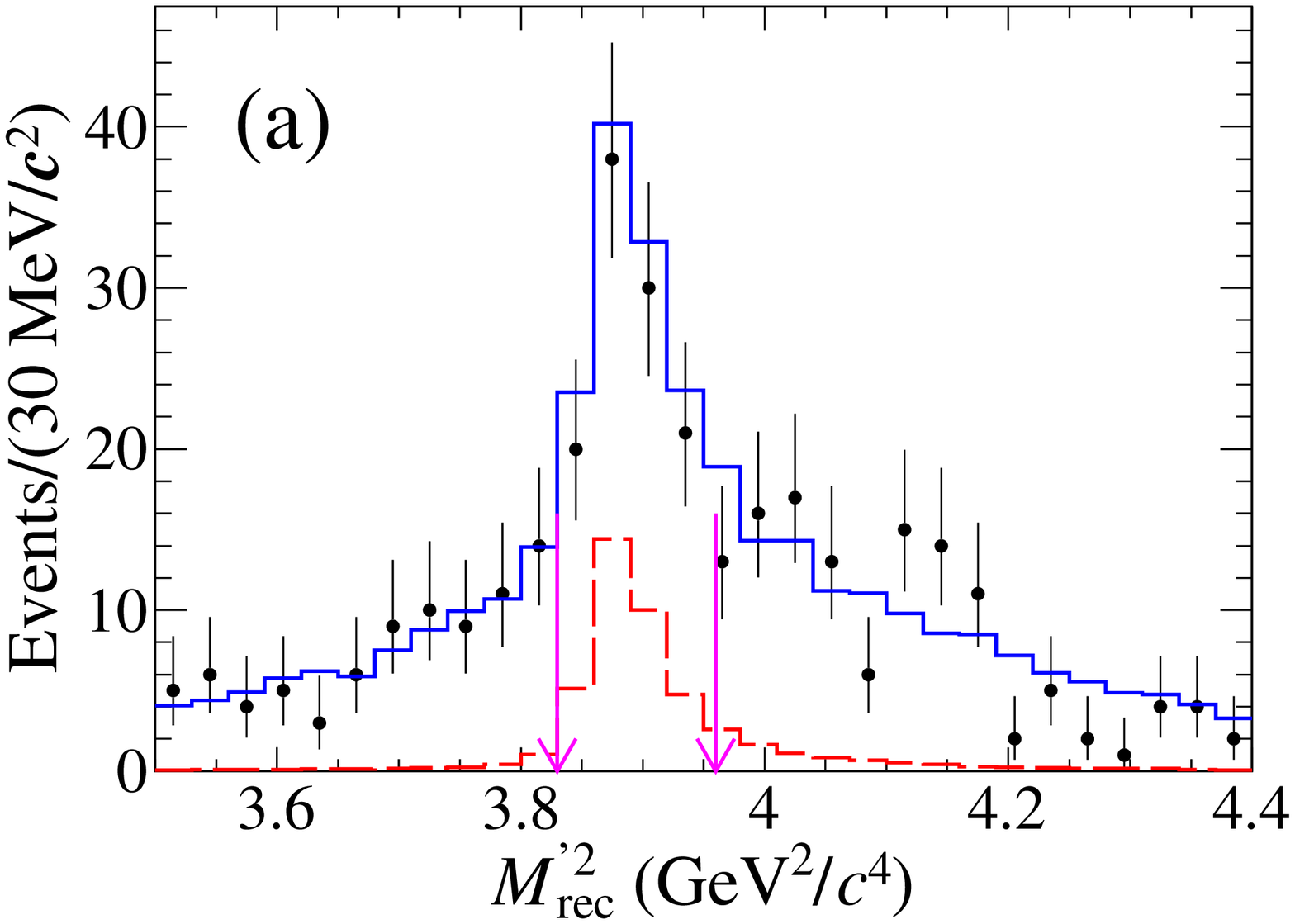}
    \includegraphics[width=0.32\textwidth]{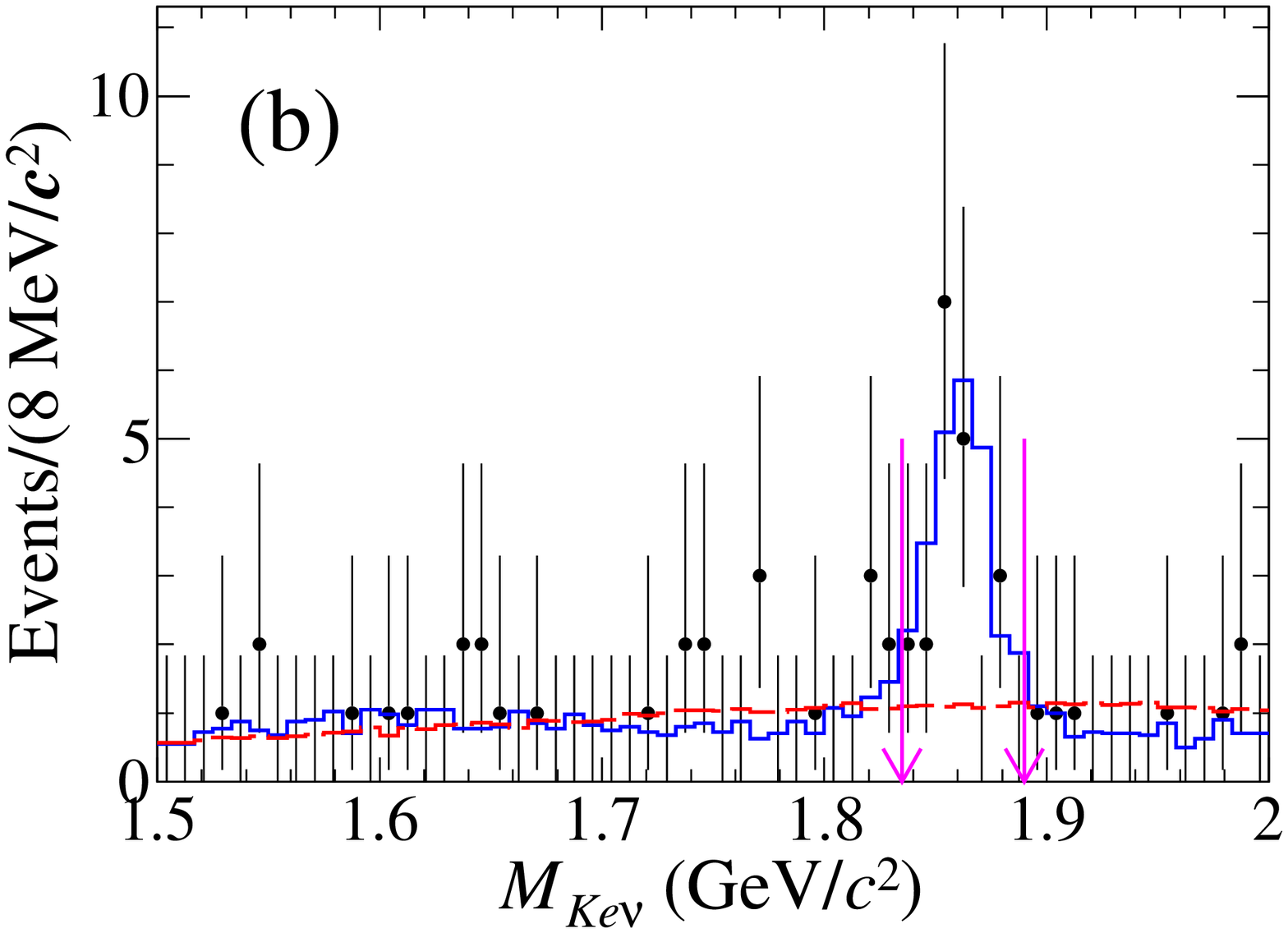}
    \includegraphics[width=0.32\textwidth]{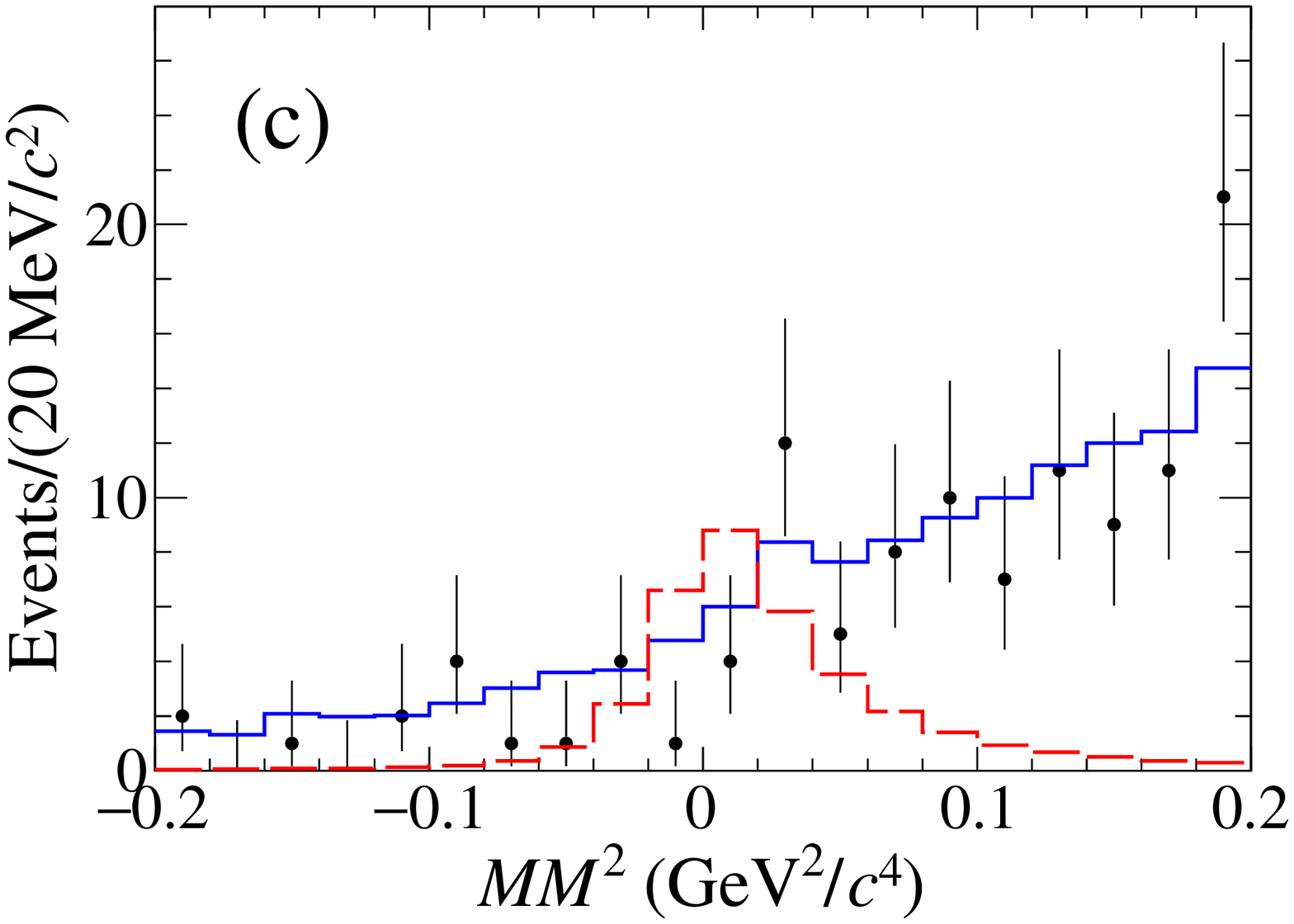}
    \caption{(a) $M^{\prime 2}_{\rm rec}$, (b) $M_{Ke\nu}$, and (c) $M\!M^2$
      distributions of data and MC samples. The points with error bars are data.
      The blue solid and red dashed lines are inclusive and signal MC samples,
      respectively. The pair of pink arrows denote the signal windows. The
      signal MC sample is normalized arbitrarily for visualization purposes. An
      additional requirement of $|M\!M^2|<0.20$~GeV$^{2}$/$c^4$ has been applied.
    }
    \label{fig:M2rec_KevTag_MM2}
  \end{center}
\end{figure*}

After a $D_s^-$ tag candidate is identified, we search for the signal
$D_s^+\to \pi^0e^+\nu_e$ candidate recoiling against the tag by requiring one
charged particle identified as $e^+$, one $\pi^0$ candidate, and at least one
more photon to reconstruct the transition photon of
$D_s^{*\pm}\to\gamma D_s^{\pm}$.  Events having charged tracks other than those
accounted for in the tagged $D_s^-$ and the electron are rejected. A kinematic
fit is performed under the hypothesis
$e^+e^-\to D_{s}^{*\pm}D_{s}^{\mp}\to \gamma D_{s}^{+}D_{s}^{-}$, with
$D_{s}^{-}$ decaying to one of the tag modes and $D_{s}^{+}$ decaying to the signal
mode. The combination with the minimum $\chi^2$ assuming a $D_s^{*+}$
meson decays to $D_{s}^{+}\gamma$ or a $D_s^{*-}$ meson decays to
$D_{s}^{-}\gamma$ is chosen. The total four-momentum is constrained
to the four-momentum of the initial $e^+e^-$ beams. Invariant masses of the tag $D_s^-$,
the signal $D_s^+$, the $D_s^*$, and $\pi^0$ are constrained to the
corresponding known masses~\cite{PDG}.
This gives us a total of eight constraints (8C). The missing neutrino four-vector needs to be
determined (-4C), so we are left with a four-constraint fit (4C). 
Furthermore, we require that the
maximum energy of photons not used in the DT event selection
is less than 0.2~GeV.
The square of the recoil mass against the transition photon and the
tag $D_s^-$~($M^{\prime 2}_{\rm rec}$) is expected to peak at the
known $D^{\pm}_s$ meson mass-squared before the kinematic fit for
signal $D^{*\pm}_s D^{\mp}_s$ events.  Therefore, we require
$M^{\prime 2}_{\rm rec}$ to satisfy $3.83 <M^{\prime 2}_{\rm
  rec}<3.96$~GeV$^{2}$/$c^4$, as shown in
Fig.~\ref{fig:M2rec_KevTag_MM2}(a).  Studies of the inclusive MC
sample show that there is a large background coming from $D^0\to K^-
e^+ \nu$ decays versus a hadronic $\bar{D}^0$ decay with a $\pi^0$
meson in the final state, where the $K^-$ and the $\pi^0$ mesons are
interchanged between the two decays. In order to remove this
background, for $D_s^-$ tag modes with a $K^-$, the invariant mass of
the final-state particles of the reconstructed tag $D_s^-$ except the
$K^-$ and the $\pi^0$ in the reconstructed signal $D_s^+$ is
calculated, called $M_{Ke\nu}$. A veto
$1.835<M_{Ke\nu}<1.890$~GeV/$c^2$ is applied as shown in
Fig.~\ref{fig:M2rec_KevTag_MM2}(b).  This veto removes more than 90\%
of this background (about 20\% of the total background) and sacrifices
only about 4\% efficiency.  The DT efficiencies are obtained using the
signal MC samples and listed in Table~\ref{tab:dtagEff2}.
\begin{table*}[htbp]
  \renewcommand\arraystretch{1.25}
  \caption{DT efficiencies ($\epsilon_{\text{tag,sig}}^{\text{DT}}$) at
    $\sqrt{s}=$ (I) 4.178~GeV, (II) 4.189-4.219~GeV, and (III) 4.226~GeV. The
    efficiencies for the energy points $4.189$-$4.219$~GeV are averaged based
    on the luminosities. The BF of the $\pi^{0}$ decay is not included.
    Uncertainties are statistical only.
  }\label{tab:dtagEff2}
  \begin{center}
    \begin{tabular}{lccc}
      \hline
      Tag mode                                              & (I) $\epsilon_{\text{tag,sig}}^{\text{DT}} (\%)$ & (II)$\epsilon_{\text{tag,sig}}^{\text{DT}} (\%)$&(III)$\epsilon_{\text{tag,sig}}^{\text{DT}} (\%)$\\
      \hline
      $D^-_s\to K^{+}K^{-}\pi^{-}$                          &  $13.94\pm0.11$        &  $13.18\pm0.06$         &  $12.20\pm0.11$ \\
      $D^-_s\to K^0_{S}K^{-}$                               &  $10.31\pm0.04$        &  $\phantom{0}9.77\pm0.02$         &  $\phantom{0}9.02\pm0.04$ \\
      $D^-_s\to K^{0}_{S}K^{-}\pi^{0}$                      &  $\phantom{0}4.78\pm 0.07$        &  $\phantom{0}4.56\pm0.03$         &  $\phantom{0}4.34\pm0.06$ \\
      $D^-_s\to K^{+}K^{-}\pi^{-}\pi^{0}$                   &  $\phantom{0}2.89\pm0.02$        &  $\phantom{0}2.79\pm0.01$         &  $\phantom{0}2.66\pm0.02$ \\
      $D^-_s\to K^0_{S}K^{-}\pi^+\pi^-$                     &  $\phantom{0}5.38\pm0.09$        &  $\phantom{0}5.03\pm0.04$         &  $\phantom{0}4.71\pm0.08$ \\
      $D^-_s\to K^0_{S}K^{+}\pi^-\pi^-$                     &  $\phantom{0}5.40\pm0.07$        &  $\phantom{0}5.15\pm0.03$         &  $\phantom{0}4.84\pm0.06$ \\
      $D^-_s\to \pi^{+}\pi^{-}\pi^{-}$                      &  $16.92\pm0.12$        &  $15.79\pm0.06$         &  $14.51\pm0.12$ \\
      $D^-_s\to \pi^{-}\eta$                                &  $13.98\pm0.14$        &  $13.02\pm0.07$         &  $12.02\pm0.13$ \\
      $D^-_s\to \pi^-\pi^0\eta$                             &  $\phantom{0}6.52\pm0.04$        &  $\phantom{0}6.07\pm0.02$         &  $\phantom{0}5.52\pm0.04$ \\
      $D^-_s\to \pi^{-}\eta^{\prime}_{\pi^{+}\pi^{-}\eta}$  &  $\phantom{0}5.60\pm0.09$        &  $\phantom{0}5.31\pm0.04$         &  $\phantom{0}4.87\pm0.09$ \\
      $D^-_s\to \pi^{-}\eta^{\prime}_{\gamma\rho^0}$        &  $\phantom{0}7.89\pm0.08$        &  $\phantom{0}7.59\pm0.04$         &  $\phantom{0}7.05\pm0.08$ \\
      $D^-_s\to K^{-}\pi^{+}\pi^{-}$                        &  $13.33\pm0.14$        &  $12.51\pm0.07$         &  $11.51\pm0.13$ \\
      \hline
    \end{tabular}
  \end{center}
\end{table*}

\begin{figure}[htp]
  \begin{center}
    \includegraphics[width=0.32\textwidth]{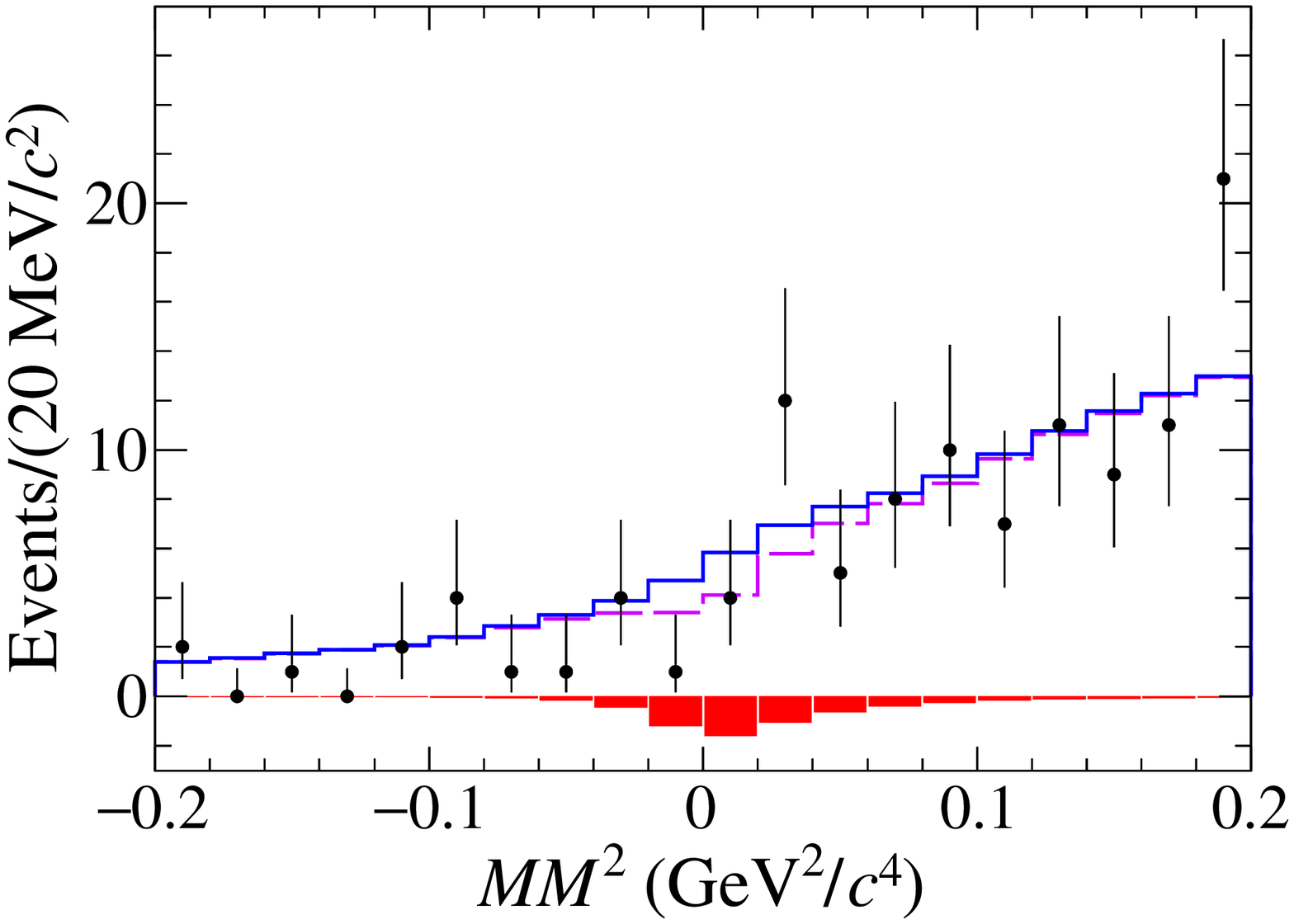}
    \caption{Fit to $M\!M^2$ distribution of data samples. The data are
      represented by points with error bars, the total fit result by the violet
      dashed line, the background by the blue solid line, and signal by the red
      filled histogram.
    }
    \label{fig:MM2fit}
  \end{center}
\end{figure}

The missing-mass squared of the neutrino is 
defined as
\begin{eqnarray}
  \begin{aligned}
    M\!M^2=\frac{1}{c^2}(p_{\rm cm}-p_{\rm tag}-p_{\pi^0}-p_{e}-p_{\gamma})^2,\,
    \label{def:MM2}
  \end{aligned}
\end{eqnarray}
where $p_{\rm cm}$ is the four-momentum of the $e^+e^-$ CM system, and
$p_{i}$ ($i=\pi^0, e, \gamma$) is the four-momentum of the final-state
particle $i$ on the signal side. The $M\!M^2$ distribution of accepted
candidate events is shown in
Fig.~\ref{fig:M2rec_KevTag_MM2}(c). Unbinned maximum-likelihood fits
to the $M\!M^2$ distribution are performed, where the signal and
background shapes are modeled by MC-simulated shapes obtained from the
signal and inclusive MC samples, respectively. The fit result is shown
in Fig.~\ref{fig:MM2fit}, and the fitted signal yield is
$-6.9\pm7.2$. Since no significant signal is observed, an upper limit
is determined with the likelihood distribution, shown in
Fig.~\ref{fig:LL_smear_data}, as a function of assumed BFs. The upper
limit on the BF at the $90\%$ confidence level, obtained by
integrating from zero to $90\%$ of the resulting curve, is
$\mathcal{B}(D^+_s\to \pi^0 e^+\nu_{e})<6.4\times 10^{-5}$.  The
method to incorporate systematic uncertainty is discussed in the next
section.

\section{SYSTEMATIC UNCERTAINTY}
\begin{figure}[htp]
  \begin{center}
    \includegraphics[width=0.40\textwidth]{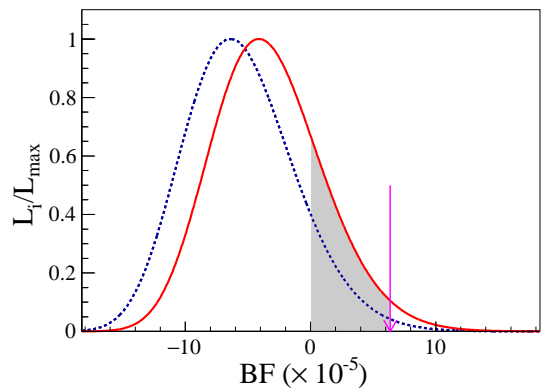}
    \caption{
      Likelihood distributions versus BF of the data samples.
      The likelihood of each bin is denoted as ${\rm L_{i}}$ and the maximum
      of the likelihood is ${\rm L_{max}}$.
      The results obtained with and without incorporating the systematic
      uncertainties are shown with red solid and blue dashed curves,
      respectively. The pink arrow shows the result corresponding to the
      90\% confidence level.
    }
    \label{fig:LL_smear_data}
  \end{center}
\end{figure}

The likelihood distribution used in the upper limit measurement covers
a range of BFs, as shown in Fig.~\ref{fig:LL_smear_data} (or signal
events yields).  The sources
of systematic uncertainties on the BF measurement are classified into
two types: additive (or independent of the measured BF central value) and multiplicative (proportional to the BF). The
multiplicative ones are summarized in
Table~\ref{tab:Bf-syst-sum}. Note that most systematic uncertainties
on the tag side cancel due to the DT technique.

Additive uncertainties affect the signal yield determination, which is
dominated by the imperfect background shape description. This
systematic uncertainty is studied by altering the nominal MC
background shape with two methods. First, alternative MC samples are
used to determine the background shape, where the relative fractions of
backgrounds from $q\bar{q}$ and non-$D_{s}^{*+}D_{s}^{-}$ open-charm
are varied within their uncertainties, and the BFs of the major
$D_s^{*}D_s$ background sources, i.e.~$D_s^+\to \eta e^+ \nu_e$,
$D_s^+\to f_0 e^+ \nu_e$, $D_s^+\to K_S^0 e^+ \nu_e$, and $D_s^+\to
\tau^+ \nu_e$, are varied by their listed
uncertainties~\cite{PDG}. Second, the background shape is obtained
from the inclusive MC samples using a Kernel estimation
method~\cite{kernel} implemented in RooFit~\cite{RooFit}. The
smoothing parameter of RooKeysPdf is varied between 0 and 2 to obtain
alternative background shapes. An alternative signal shape based on
the simple pole model~\cite{Becher:2005bg}
is tested, but the associated uncertainty is
negligible.

Multiplicative uncertainties are from the efficiency determination and the
quoted BFs. The uncertainties in the total number of the ST $D_s^-$ mesons is
assigned to be 0.5\% by examining the changes of the fit yields when varying
the signal shape, background shape, and taking into account the background
fluctuation in the fit. The uncertainty from the BFs of $D^*_s\to \gamma D_s$
and $\pi^0\to\gamma\gamma$ decays are $0.7\%$ and $0.03\%$, respectively,
according to the known values~\cite{PDG}. The systematic uncertainty
related to $e^+$ tracking or PID efficiency is assigned as 1.0\% from studies
of a control sample of radiative Bhabha events. The systematic uncertainties associated with
reconstruction efficiencies of the transition photon and $\pi^0$ are studied by
using control samples of the decay $J/\psi\to\pi^{+}\pi^{-}\pi^{0}$ and the
process $e^+e^-\to K^+K^-\pi^+\pi^-\pi^0$, respectively. The efficiency
difference between data and MC samples is then determined to be 1.0\% for the
transition photon and 2.0\% for the final state $\pi^0$. The uncertainties due to
the maximum energy of photons not used in the DT event selection criteria, and requiring one charged track are assigned as $0.5\%$ and
$0.9\%$, respectively. We determine these uncertainties by analyzing DT hadronic
events in which one $D_s^{-}$ decays into one of the tag modes and the other
$D_s^{-}$ decays into
$K^+K^-\pi^{-}$ or $K_{S}^{0}K^{-}$. The uncertainty due to the limited MC
sample size is obtained by
$\sqrt{\sum_{\alpha}{(f_\alpha \frac{\delta\epsilon_\alpha}{\epsilon_\alpha})^2}}\approx 0.5\%$,
where $f_{\alpha}$ is the tag yield fraction, and $\epsilon_{\alpha}$ and
$\delta_{\epsilon_{\alpha}}$ are the signal efficiency and the corresponding
uncertainty of tag mode $\alpha$, respectively.
The acceptance efficiencies of the kinematic fit
requirement are studied with the control sample $D_{s}^{+}\to \pi^+\pi^0\eta$
from the DT hadronic $D_s^-D_s^{*+}+c.c.$ events due to its similar topology
and large BF. We take into account the difference of the acceptance
between data and MC simulation and the statistical uncertainty of this control
sample, and assign 0.8\% as the corresponding uncertainty.
We test an alternative simple pole model in place of the ISGW2 model in generating the
signal MC sample for the determination of detection efficiency. The form factor
of simple pole model is defined as
$f_{+}^{q^{2}}=\frac{1}{1-\frac{q^2}{M_{\rm pole}}}$, where $q$ is the four
momentum transfer and the pole mass $M_{\rm pole}$ is the known $D_s^*$
mass~\cite{PDG}. The difference of the signal efficiencies between the
two models
is assigned as the systematic uncertainties related to the MC model. The
uncertainty associated with the ST efficiency in Eq.~(\ref{eq:Bsig-gen}) is
not canceled fully, which results in a so called ``tag bias'' uncertainty,
because the ST efficiencies estimated with the generic and the signal MC
samples differ slightly due to the different multiplicities. We study the
tracking/PID efficiencies in different multiplicities, and take the combined
differences between data and MC simulation, 0.4\%, as the corresponding
uncertainty. The BF of $\pi^0\to\gamma\gamma$ is
$(98.823\pm0.034)\%$~\cite{PDG}, which causes a negligible uncertainty, 0.03\%.
By adding these uncertainties in quadrature, the total multiplicative
systematic uncertainty $\sigma_{\epsilon}$ is estimated to be 3.9\%.

\begin{table}[htp]
 \centering
 \caption{Multiplicative systematic uncertainties. All the uncertainties are relative and given in \%.}
\begin{tabular}{l|c}
\hline
Source                                 & $\sigma_{\epsilon}$ (\%) \\
\hline
$D_{s}^{-}$ yield                      & 0.5\\
${\cal B}(D^*_s\to\gamma D_s)$         & 0.7\\
$e^+$ tracking efficiency              & 1.0\\
$e^+$ PID efficiency                   & 1.0\\
$\gamma$ and $\pi^0$ reconstruction    & 3.0\\
Energy of extra photon                 & 0.5\\
No extra track                         & 0.9\\
MC statistics                          & 0.5\\
Kinematic fit                          & 0.8\\
Signal model                           & 0.9\\
Tag bias                               & 0.4\\
\hline
Total                                  & 3.9\\
\hline
\end{tabular}
\label{tab:Bf-syst-sum}
\end{table}

To take into account the additive systematic uncertainty, the
maximum-likelihood fits are repeated using different alternative
background shapes as mentioned in the previous Section and the one
resulting in the most conservative upper limit is chosen.  Finally, the
multiplicative systematic uncertainty $\sigma_{\epsilon}$ is incorporated in the
calculation of the upper limit via~\cite{Stenson, CPC-39-103001}
\begin{eqnarray}
  \begin{aligned}
    L\left(\mathcal{B}\right)\propto \int^1_0 L\left(\mathcal{B}\frac{\epsilon}{\epsilon_{0}}\right){\rm exp}\left[\frac{-\left(\epsilon/\epsilon_0-1\right)^2}{2 (\sigma_{\epsilon})^2 }\right]d\epsilon \,,
  \end{aligned}
\end{eqnarray}
where $L(\mathcal{B})$ is the likelihood distribution as a function of assumed
BFs; $\epsilon$ is the expected efficiency and $\epsilon_0$ is the averaged
MC-estimated efficiency. The likelihood distributions with and without
incorporating the systematic uncertainties are shown in Fig.~\ref{fig:LL_smear_data}.

\section{Conclusion} \label{CONLUSION}
Using a data sample corresponding to an integrated luminosity of
$6.32~\mathrm{fb}^{-1}$, taken at $\sqrt{s} = $ 4.178-4.226~GeV recorded by
the BESIII detector, we perform the first search for $D_s^+\to \pi^0 e^+\nu_e$.
No significant signal of the semileptonic decay
$D_s^+\to \pi^0 e^+\nu_e$ is observed. We set an upper limit on
$\mathcal{B}(D^+_s\to \pi^0 e^+\nu_{e})<6.4\times 10^{-5}$ at the 90\%
confidence level. Our result
is consistent with the predicted BF of $D_s^+\to\pi^0 e^+\nu_e$,
$(2.65\pm0.38)\times 10^{-5}$~\cite{plb-811-135879}, based on the mechanism of $\pi^0$-$\eta$
mixing.

\begin{acknowledgements}
\label{sec:acknowledgement}
\vspace{-0.4cm}
The BESIII collaboration thanks the staff of BEPCII and the IHEP computing center for their strong support. This work is supported in part by National Key Research and Development Program of China under Contracts Nos. 2020YFA0406400, 2020YFA0406300; National Natural Science Foundation of China (NSFC) under Contracts Nos. 11625523, 11635010, 11735014, 11822506, 11835012, 11875054, 11935015, 11935016, 11935018, 11961141012, 12192260, 12192261, 12192262, 12192263, 12192264, 12192265; the Chinese Academy of Sciences (CAS) Large-Scale Scientific Facility Program; Joint Large-Scale Scientific Facility Funds of the NSFC and CAS under Contracts Nos. U2032104, U1732263, U1832207; CAS Key Research Program of Frontier Sciences under Contracts Nos. QYZDJ-SSW-SLH003, QYZDJ-SSW-SLH040; 100 Talents Program of CAS; INPAC and Shanghai Key Laboratory for Particle Physics and Cosmology; ERC under Contract No. 758462; European Union Horizon 2020 research and innovation programme under Contract No. Marie Sklodowska-Curie grant agreement No 894790; German Research Foundation DFG under Contracts Nos. 443159800, Collaborative Research Center CRC 1044, FOR 2359, GRK 2149; Istituto Nazionale di Fisica Nucleare, Italy; Ministry of Development of Turkey under Contract No. DPT2006K-120470; National Science and Technology fund; Olle Engkvist Foundation under Contract No. 200-0605; STFC (United Kingdom); The Knut and Alice Wallenberg Foundation (Sweden) under Contract No. 2016.0157; The Royal Society, UK under Contracts Nos. DH140054, DH160214; The Swedish Research Council; U. S. Department of Energy under Contracts Nos. DE-FG02-05ER41374, DE-SC-0012069.
\end{acknowledgements}

\end{document}